\documentclass[aps, pre, reprint, groupedaddress, nofootinbib, showpacs, showkeys]{revtex4-1}
\usepackage{newfloat}
\usepackage[noend]{algcompatible}
\usepackage[size=small]{caption}
\usepackage{newfloat}
\DeclareFloatingEnvironment[
     fileext=loa,
     listname=List of Algorithms,
     name=ALGORITHM,
     placement=tbhp,
]{algorithm}
\DeclareCaptionFormat{algorithms}{\vskip-15pt\hrulefill\par#1#2#3\vskip-6pt\hrulefill}
\usepackage{amsmath, amsthm, amssymb, amsfonts, rotating}    % need for subequations
\usepackage[title,titletoc,toc]{appendix}
\usepackage{graphicx}   % need for figures
\usepackage{verbatim}   % useful for program listings
\usepackage{color}      % use if color is used in text
\usepackage{latexsym, subfig, overpic}
\usepackage[stable]{footmisc}

\newtheorem{thm}{Theorem}

\theoremstyle{definition}

\newcommand{\energy}{{\bf \varepsilon}}
\newcommand{\E}{\mathbb{E}}

\newcommand{\C}{\mathcal{C}}

\newcommand{\al}{\alpha}
\newcommand{\la}{\lambda}
\newcommand{\hidden}[1]{}
\newcommand{\F}{\mathbb{F}}
\renewcommand{\path}[2]{\mathcal{P}_{{#1},{#2}}}

\DeclareMathOperator{\allk}{\mathcal{A}_{\text{\sc{cycle}}}}
\DeclareMathOperator{\bs}{\mathcal{A}_{\text{\sc{edge}}}}

\DeclareMathOperator{\ODD}{o}

\DeclareMathOperator{\current}{1}
\DeclareMathOperator{\children}{2}
\DeclareMathOperator{\Boltz}{\mathcal{B}}

\begin{document}

\title[Stratified Sampling for the Ising Model: A Graph-Theoretic 
Approach]{Stratified Sampling for the Ising Model: A Graph-Theoretic 
Approach} 
\thanks{Official contribution of the National Institute of Standards and Technology; 
not subject to copyright in the United States.  Certain commercial products may 
be identified in order to adequately specify or describe the subject matter of 
this work. In no case does such identification imply recommendation or
endorsement by the National Institute of Standards and Technology, nor
does it imply that the equipment identified is necessarily the best
available for the purpose.}
\author{Amanda Streib}
\email{amanda.streib@nist.gov}
\affiliation{National Institute of Standards and Technology}
\author{Noah Streib}
\email{noah.streib@nist.gov}
\affiliation{National Institute of Standards and Technology}
\author{Isabel Beichl}
\email{isabel.beichl@nist.gov}
\affiliation{National Institute of Standards and Technology}
\author{Francis Sullivan}
\email{fran@super.org}
\affiliation{Center for Computing Sciences}

\date{\today}

\begin{abstract}
We present a new approach to a classical problem in statistical 
physics: estimating the partition function and other thermodynamic 
quantities of the ferromagnetic Ising model. Markov chain Monte Carlo 
methods for this problem have been well-studied, although an 
algorithm that is truly practical remains elusive.  Our approach takes 
advantage of the fact that, for a fixed bond strength, studying the 
ferromagnetic Ising model is a question of counting particular subgraphs 
of a given graph.  We combine graph theory and heuristic sampling to 
determine coefficients that are independent of temperature 
and that, once obtained, can be used to determine the partition function 
and to compute physical quantities such as mean energy, mean magnetic moment, 
specific heat, and magnetic susceptibility.
\end{abstract}



\maketitle

\section{Introduction}
Computing thermodynamic quantities of the ferromagnetic
Ising model has been a fundamental problem in statistical physics 
since the early 20th century \cite{Ising}, where the demonstration 
of the model's phase transition served as the first rigorous proof 
that small changes at an atomic scale can lead to
large, observable changes~\cite{Peierls}.  Singularities in the 
thermodynamic quantities indicate the critical temperature at which 
the phase transition occurs. The partition function $Z$ of
the Ising model and its partial derivatives determine these quantities.
While $Z$ has been found exactly in special cases~\cite{Onsager, Yang}, 
there is unlikely to exist an efficient
method of finding $Z$ in general \cite{JS}.  Therefore, the task of 
estimating $Z$ has drawn significant
effort from the physics and computer science communities~\cite{cipra}.
However, an algorithm that is truly practical has yet to be found.
In this paper, we present a new heuristic sampling approach with the
goal of solving real-world instances quickly.

The classical approach to this problem is to sample from the Gibbs
distribution using a Markov chain~\cite{Metropolis, SW, JS}. Ideally, the 
algorithm will require only a polynomial number of samples to estimate 
$Z$ at a particular temperature, but even then this process must be 
repeated for each temperature of interest.  In contrast, each
run of our heuristic sampling algorithm, $\allk$, estimates certain 
coefficients that are independent of temperature.  Once obtained, 
these coefficients can be used to compute $Z$, mean energy, mean 
magnetization, specific heat, and magnetic susceptibility at all 
temperatures by simply evaluating polynomials with these coefficients.

For a fixed bond strength, computing $Z$ 
is equivalent to counting subgraphs of a graph $G$.  Let $x_{k,e}$ 
denote the number of subgraphs of $G$ with $2k$ odd vertices and $e$ 
edges.  Using the high-temperature expansion, we can write $Z$ and 
its derivatives as polynomials whose coefficients come from the set 
of $x_{k,e}$. For each $k$, $\allk$ generates a search tree whose 
leaves are the set of subgraphs with $2k$ odd vertices, and then 
implements the stratified sampling method of Chen~\cite{Chen} to 
estimate the $x_{k,e}$. In the absence of an applied field, the 
problem of estimating $Z$ reduces to estimating $x_{0,e}$ for all $e$.
As will become clear, it is simple to restrict $\allk$ to subgraphs 
with no odd-degree vertices, which significantly reduces the complexity 
of the algorithm in this special case.

\section{Definitions and Terminology}\label{s:term}
In this section, we introduce important notions from 
statistical physics and graph theory.  

\subsection{Ising Model}
Given a graph $G=(V,E)$ with $|V|=n$ and $|E|=m$, a \emph{spin 
configuration} $\sigma =\sigma(G)$ is an assignment of spins in 
$\{+ 1, -1\}$ to the elements of $V$. The energy of $\sigma$ is given 
by the Hamiltonian
$$H(\sigma) = -J\sum_{(x,y)\in E}\sigma_x\sigma_y - B \sum_{x\in V} 
\sigma_x,$$
where $J$ is the interaction energy (bond strength) and $B$ is the 
external magnetic field. In this paper we restrict the ferromagnetic case, 
fixing $J=1$.To model the physical reality of a ferromagnet,
the probability assigned to state $\sigma$ is given by the Gibbs 
distribution, defined as $e^{-\beta H(\sigma)}/Z$,
where $\beta = (k_{\Boltz}T)^{-1}$ is proportional to inverse 
temperature and $k_{\Boltz}$ is Boltzmann's constant.
The normalizing constant $Z = \sum_{\sigma} \exp(-\beta H(\sigma))$
is also called the partition function.

Following the notation of \cite{JS}, let $\lambda
= \tanh(\beta J)$ and $\mu=\tanh(\beta B)$.  The high-temperature
expansion is defined by $Z=AZ'$, where $A=(2\cosh(\beta B))^n 
\cosh(\beta J)^m$
is an easily computed constant, and
\begin{equation*}\label{htexp}
Z'= \sum_{X\subseteq E} \lambda^{|E(X)|}\mu^{|\ODD(X)|}~,
\end{equation*}
where the sum is taken over all subsets $X$ of the edges of $G$.  In a 
slight abuse of notation, we let $X$ also refer to the graph with 
vertex-set $V$ and edge-set $X$.  In this manner, $E(X)$ is the edge-set 
of $X$, $\ODD(X)$ is the the set of odd-degree vertices in $X$, and 
all subgraphs in this paper are spanning and labeled.

Since all graphs have an even number of vertices of odd degree,
Jerrum and Sinclair \cite{JS} write $Z'$ as a polynomial in $\mu^2$:
$ Z' = \sum_{k=0}^{\lfloor n/2 \rfloor} c_k \mu^{2k},$ where $ c_k = 
\sum_{X~:~|\ODD(X)|=2k}
\la^{|E(X)|}~.$
Notice that we can compute $Z'$ for any choice of $\mu$ given the values 
of the $c_k$,
making the $c_k$ independent of the magnetic field.  However, we wish
to have full temperature-independence, so we write 
\begin{equation}\label{eq:c_k}
c_k = \sum_{e=0}^m x_{k,e}\lambda^e \quad \text{and} \quad Z' =
\sum_{k=0}^{\lfloor n/2 \rfloor} \sum_{e=0}^m x_{k,e}\lambda^e
\mu^{2k},
\end{equation}
where $x_{k,e}$ is as defined in the introduction.
As we shall see, $\allk$ is designed to estimate the $x_{k,e}$.
Thus, $\allk$ yields an estimate of $Z'$, and hence $Z$ as well, at all 
temperatures simultaneously.

While $\allk$ is defined for all graphs $G$, the graphs with the most 
physical significance are the square lattices (grids) with periodic 
boundary conditions in two and three dimensions.
Therefore, all of the computations provided in this paper utilize such 
graphs, and we shall refer to the $s \times s$ square lattice with 
periodic boundary conditions simply as the $s \times s$ grid.

\subsection{Cycle Bases}\label{s:cycle_bases_def}
We now introduce some elementary algebraic graph theory
which $\allk$ uses (for more on this topic, see \cite{Diestel}).  
The \emph{symmetric difference} of two 
subgraphs $X_1$ and $X_2$ of $G$, written $X_1 \oplus X_2$, is the 
subgraph of $G$ that contains precisely those edges in exactly one of 
$X_1$ and $X_2$. One may consider this operation as addition of 
subgraphs over the field $\F_2 = \{0,1\}$. 
Notice that an edge $e$ is in $\bigoplus_{i=1}^t X_i$ if and only 
if $e$ appears in an odd number of these subgraphs.

Let $\mathcal{E}_0$ be the set of \emph{even subgraphs}, those 
subgraphs with no vertices of odd degree. Since the symmetric 
difference of two even subgraphs is again an even subgraph, 
we may view $\mathcal{E}_0$ as a vector space over $\F_2$, 
called the \emph{cycle space} of $G$. The dimension of the cycle 
space is $m-n+1$. Hence, every set of $m-n+1$ linearly independent 
even subgraphs forms a \emph{cycle basis} $\mathcal{C}$ of $G$. 
Further, every even subgraph has a \emph{unique} 
representation using the elements of $\mathcal{C}$, and 
$|\mathcal{E}_0| = 2^{m-n+1}$.

When $X \in \mathcal{E}_0$, the parity of each vertex in $X \oplus Y$ 
is the same in $Y$.  Now consider a subgraph $P$ of $G$ with 
$\ODD(P)=\{v_1,v_2,\ldots, v_{2k}\}$. The set
 $$\mathcal{E}_0 \oplus P := \{X\oplus P: X\in \mathcal{E}_0\}$$
is exactly the $2^{m-n+1}$ subgraphs whose odd vertices are 
$\ODD(P)$.  Therefore, the set of subgraphs with $2k$ odd vertices, 
$\mathcal{E}_k$, is $\bigcup_{S} \mathcal{E}_0\oplus P_S$, where 
the union is over all $S \subseteq V$ of size $2k$ and $P_S$ is 
\emph{any} subgraph with $\ODD(P)=S$.

Cycle bases have a long history in combinatorics \cite{maclane},
and are used both in theory and applications 
\cite{cycle_basis_survey}.
A \emph{fundamental cycle basis} is defined as the cycles 
in $T+e$ for each $e \in E(G)-E(T)$, for a spanning tree $T$ of $G$.  
Since spanning trees can be found quickly (see e.g. \cite{CLRS}), 
so can fundamental bases.   
\emph{Minimum cycle bases}, which are bases with the
fewest total edges, have proven helpful in practice
and can also be found in 
polynomial time \cite{min_cycle_basis}.

\section{Algorithms}\label{s:algs}
Our main data structure is a search-tree; a rooted tree in which each
node represents a subgraph of $G$.  For each $k$, we shall define a
search-tree $\tau_k$ whose leaves are precisely $\mathcal{E}_k$.
Our goal is to estimate $x_{k,e}$, the number of leaves of $\tau_k$
that have $e$ edges.

Tree search algorithms have a lengthy history in computer science \cite{Pearl}.  
A classical example of such is an algorithm of Knuth
\cite{Knuth} for estimating properties of a backtrack tree.  To estimate
the number of leaves, for example, Knuth's algorithm explores a random
path down the tree from the root, choosing a child uniformly at
random at each step.  It then returns the product of the number of
children of each node seen along the path.  It is easy to see that this
estimator is unbiased; i.e. the expected value is the number of leaves.

For our application, we want the number
of leaves of $\tau_k$ of a certain type (with $e$ edges).  We achieve this via
Chen's generalization of Knuth's algorithm, which was originally
introduced to reduce the variance of the estimator. 
Since Chen's work lies at the heart of our approach, we take the next section to
explain it in further detail.  In Section~\ref{s:allk}, we describe $\allk$.  
In \cite{SM}, we present an alternative to $\allk$, 
which is related to~\cite{JS}.  This approach, which we call $\bs$, 
may be more appropriate in the presence of an external field, 
but is outperformed by $\allk$ when $B=0$. 

\subsection{Stratified Sampling}
We describe in Algorithm \ref{alg:Chen} a simplified version of the 
stratified sampling algorithm introduced by Chen \cite{Chen}.  
Let $\tau$ be a search tree and choose a \emph{stratifier} for $\tau$
--- a way of partitioning the nodes into sets called \emph{strata}.\footnote{In 
general, the stratifier must satisfy a few technical conditions.  \
However, as long as we require each strata to contain nodes from a 
single level of $\tau$, we are guaranteed that these conditions are met.}  
For each stratum $\alpha$, Algorithm \ref{alg:Chen}
produces a representative $s_{\al} \in \al$ and a weight
$w_{\al}$, which is an unbiased estimate of the number of nodes in $\al$. 

For Algorithm \ref{alg:Chen}, let $Q_{\current}$ and $Q_{\children}$ be queues.  
Each node $s$ of $\tau$ has a weight $w$, and we write $(s,w)$ 
to represent this pair.  The input is the root $r$ of $\tau$, a method
for determining the children of a node in $\tau$, and the stratifier.
The output is the set of $(s_\al,w_\al)$.  If the algorithm never
encounters an element of $\al$, it returns $(\emptyset, 0)$ for $\al$.
 
\begin{algorithm}
\caption{Chen's Algorithm}
\begin{algorithmic}
\STATE \emph{initialize}: $Q_{\current}=\{(r,1)\}$, $Q_{\children}=\{\}$, $i =0$.
\WHILE{$i < \text{number of levels in $\tau$}$}
\WHILE{$Q_{\current} \not= \emptyset$}
 \STATE output the first element $(s,w)$ of $Q_{\current}$
  \FOR{each child $t$ of $s$ in $\tau$}
   \IF{$Q_{\children}$ contains an element $(u,w_u)$ in the same stratum as $t$}
    \STATE update $w_u = w + w_u$
    \STATE w. prob. $w/w_u$ replace $(u,w_u)$ with $(t,w_u)$ in $Q_{\children}$
   \ELSE
     \STATE add $(t,w)$ to $Q_{\children}$
   \ENDIF
  \ENDFOR
  \STATE pop $(s,w)$ off of $Q_{\current}$
 \ENDWHILE
 \STATE set $Q_{\current} = Q_{\children}$ and reset $Q_{\children}=\emptyset$
 \STATE $i$++
\ENDWHILE
\end{algorithmic}
\noindent\hrulefill\par\nobreak\vskip-5pt
\label{alg:Chen}
\end{algorithm}

\subsection{Cycle-Addition Algorithm}\label{s:allk}
Let $S \subseteq V$ of size $2k$ and recall that $P_S$ is any subgraph
of $G$ with $\ODD(P) = S$.  Let $\C = \{C_1, C_2,\ldots, C_{m-n+1}\}$ be
a cycle basis of $G$.  Define $\tau(\C,P_S)$ as the search-tree
determined by the following rules:
\begin{itemize}
\item[1.] $P_S$ is the root of $\tau(\C,P_S)$, and
\item[2.] each node $X$ at level $0 \leq i < m-n+1$ has two children: 
$X \oplus \mathcal{C}_{i-1}$ and $X$.
\end{itemize}
Now $\tau_k$ is the tree with artificial root node $R$ whose 
$\binom{n}{2k}$ children correspond to the roots of $\tau(\C,P_S)$,
one for each distinct subset of size $2k$.

\begin{figure}[!htb]
\centering
\begin{overpic}[scale=.5]{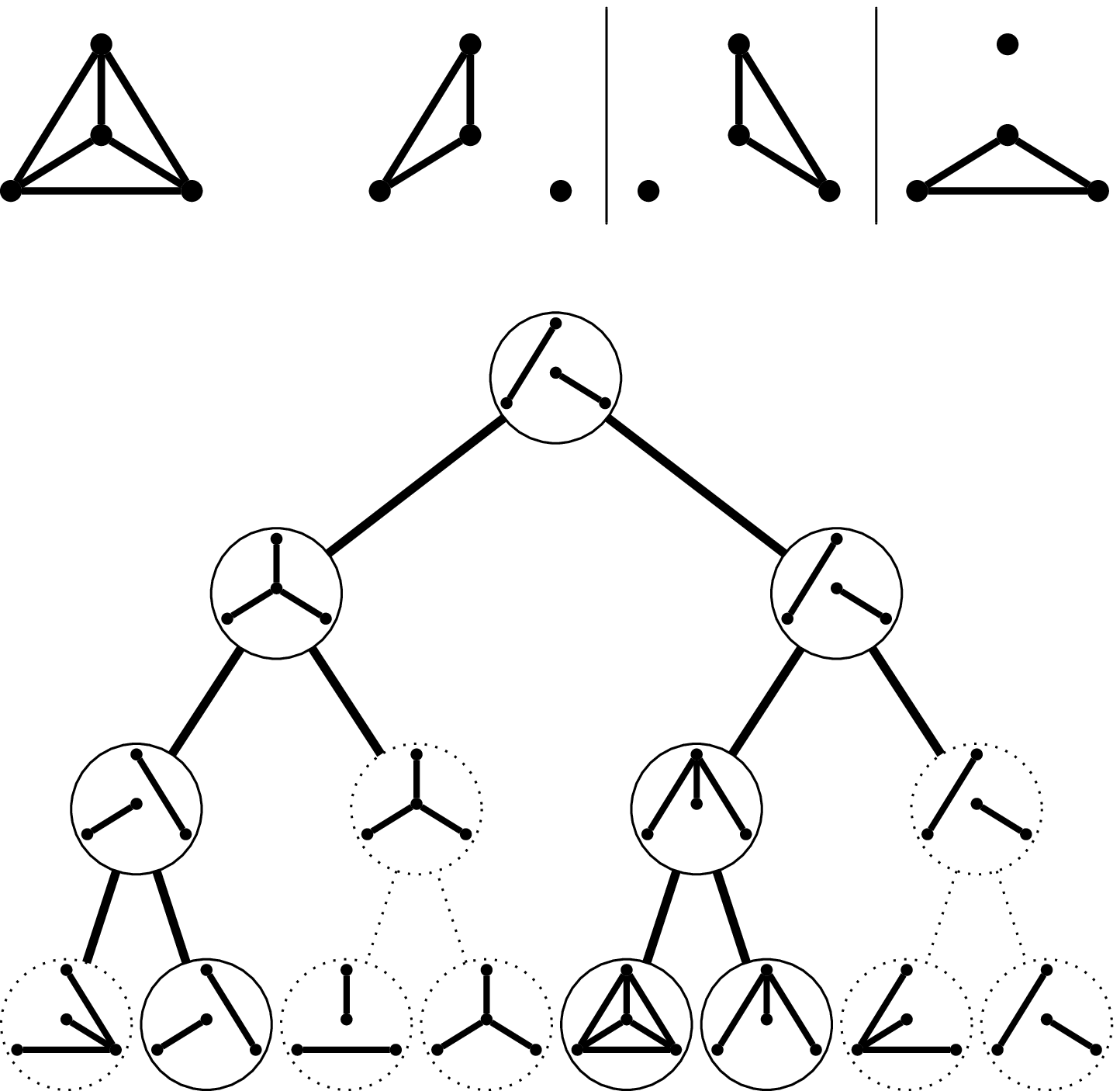}
\put(8, 75){$G$} \put(40.5, 75){$C_1$}
\put(64.5, 75){$C_2$} \put(88, 75){$C_3$}
\put(38.5, 66){$P_S$} %\put(5, 58){$\tau(\mathcal{C}, P_S)$}
\put(28, 55){$\oplus C_1$}
\put(9, 35){$\oplus C_2$} \put(59, 35){$\oplus C_2$}
\put(-.5, 14.5){$\oplus C_3$} \put(24.75, 14.5){$\oplus C_3$}
\put(49.75, 14.5){$\oplus C_3$} \put(75.25, 14.5){$\oplus C_3$}
\put(53.5, 69){${\bf 1}$}
\put(28.25, 49.5){${\bf 1}$} \put(79, 49.5){${\bf 1}$}
\put(17, 29){${\bf 2}$} \put(67.5, 29){${\bf 2}$}
\put(22, 11){${\bf 2}$} \put(60, 11){${\bf 2}$} \put(72, 11){${\bf 4}$}
\end{overpic}
\caption{}
\label{fig:allk}
\end{figure}

In order to implement Algorithm~\ref{alg:Chen}, we define the
stratifier for each $\tau(\C,P_S)$ by: the nodes $X$ and $Y$ in
$\tau(\C,P_S)$ belong to the same stratum if and only if
$X$ and $Y$ are in the same level of $\tau(\C,P_S)$, and 
$|E(X)| = |E(Y)|$.

The inputs to $\allk$ are a graph
$G$, an integer $k$ in $[0, n/2]$,
and an integer $N$. The output of
each of the $N$ runs of Algorithm \ref{alg:Chen}, as a subroutine of
$\allk$, is a set of $(s_{\al},w_{\al})$ pairs.
Consider a representative $s_{\al}$ that is a leaf node in the tree
$\tau(\C,P_S)$, and suppose $s_{\al}$ has $e$ edges.  Then
$\binom{n}{2k}w_{\al}$ is our estimate of $x_{k,e}$, 
since each sample represents all $\binom{n}{2k}$ choices of $S$.  

\begin{algorithm}
\caption{$\allk$}
\begin{algorithmic}
\STATE Choose a cycle basis $\C$ of $G$
\FOR{$j \in [1,N]$}
  \STATE Choose $S \subseteq V$ with $|S| = 2k$
  \STATE Find $P_S$
  \STATE Run Algorithm \ref{alg:Chen} on $\tau(\C,P_S)$
\ENDFOR
\FOR{$e \in [0,m]$}
  \STATE Let $\al$ be the stratum corresponding to the bottom level of
  $\tau(\C,P_S)$ and $e$ edges, and output $\binom{n}{2k}$ times the 
  average of the $N$ estimates of $w_\al$ as $x_{k,e}$  
\ENDFOR
\end{algorithmic}
\noindent\hrulefill\par\nobreak\vskip-5pt
\label{alg:allk}
\end{algorithm}

Figure \ref{fig:allk} shows an example of $\allk$ with $k=2$, $S=V(G)$, 
$N=1$, and $G$, $P_S$, $\mathcal{C}=\{C_1,C_2,C_3\}$, and 
$\tau(\mathcal{C}, P_S)$ as depicted.  The graphs bounded by solid circles
are the strata representatives, and their weights are in bold
just above.  The solid edges of $\tau(\mathcal{C}, P_S)$ connect the 
nodes seen by $\allk$.
The output is $x_{2,2} = 2$, $x_{2,3} = 4$, $x_{2,6}=2$, and 
$x_{2,e}=0$ for $e \in \{0,1,4,5\}$.

\begin{figure}[!htb]
\centering
\subfloat[]{
\begin{overpic}[scale=.19]{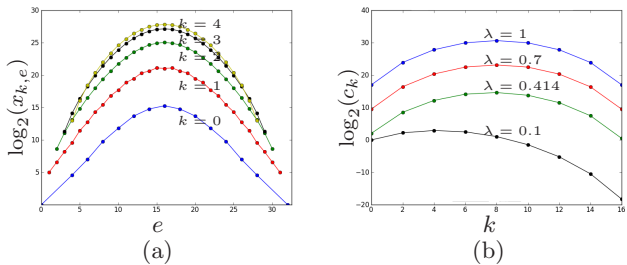}
\put(55,32){\tiny{$k=0$}} \put(55,43){\tiny{$k=1$}}
\put(55,52){\tiny{$k=2$}} \put(55,57){\tiny{$k=3$}}
\put(55,62){\tiny{$k=4$}} 
\put(2,20){\begin{sideways}\parbox{15mm}{\footnotesize{$\log_2(x_{k,e})$}}
\end{sideways}}
\put(47,-1){\small{$e$}}
\end{overpic}\label{fig:xkes_4x4x1}}
\subfloat[]{
\begin{overpic}[scale=.19]{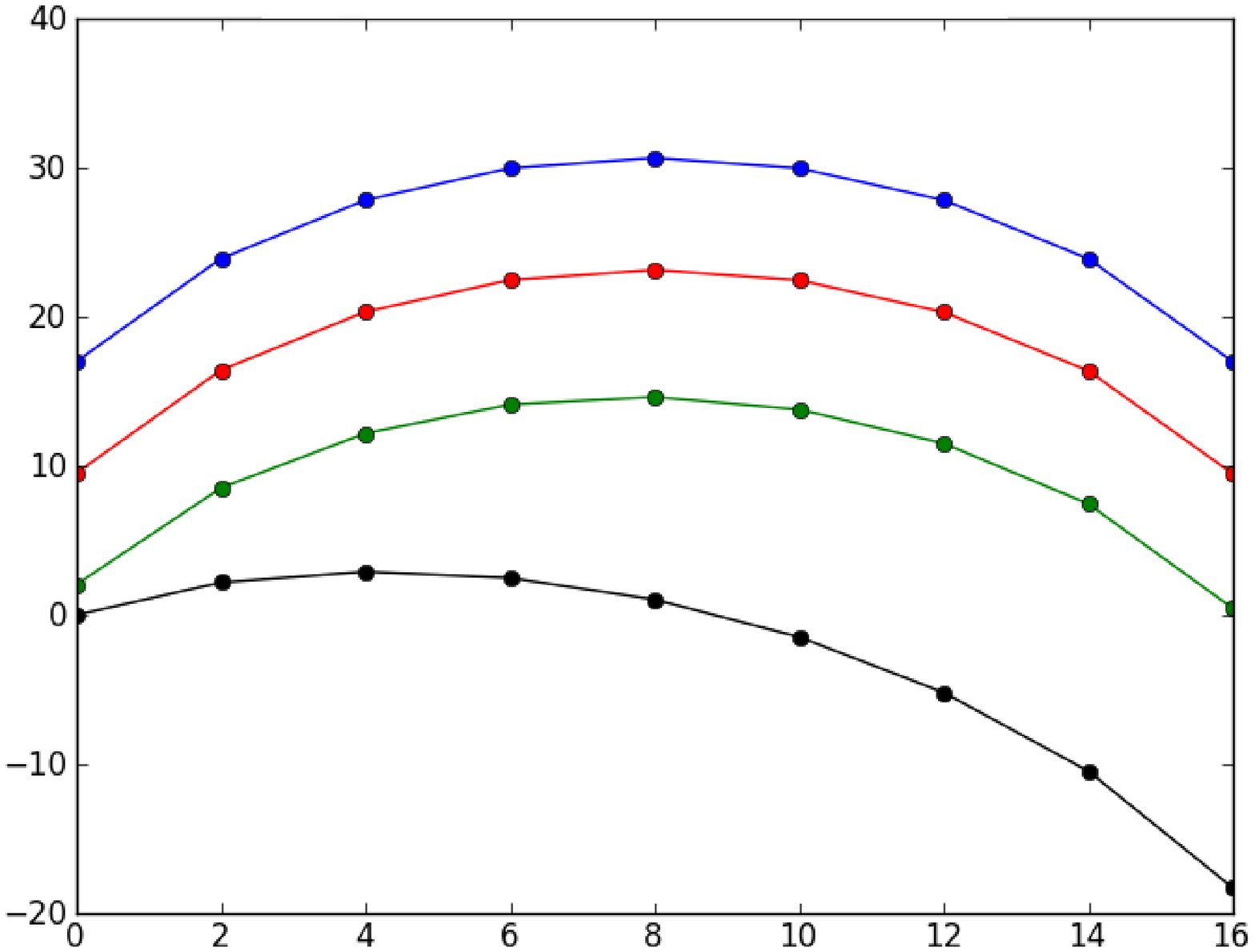}
\put(47,59){\tiny{$\lambda=1$}} \put(47,51){\tiny{$\lambda=0.7$}}
\put(47,43){\tiny{$\lambda=0.414$}} \put(47,29){\tiny{$\lambda=0.1$}}
%\put(73,55){\tiny{$\lambda=1$}} \put(73,47){\tiny{$\lambda=0.7$}}
%\put(73,39){\tiny{$\lambda=0.414$}} \put(73,20){\tiny{$\lambda=0.1$}}
\put(2,20){\begin{sideways}\parbox{15mm}{\footnotesize{$\log_2(c_k)$}}
\end{sideways}}
\put(47,-1){\small{$k$}}
\end{overpic}\label{fig:cks_4x4x1}}
\caption{(Color online)}
\label{fig:xkes}
\end{figure}

In Figure~\ref{fig:xkes_4x4x1}, we show the output of many runs of $\allk$ 
on a $4\times 4$ grid for $k \in [0,4]$, and use this output (and that for 
$k \in [5,8]$) with Equation~\ref{eq:c_k} to get Figure~\ref{fig:xkes_4x4x1}, 
using four values of $\lambda$.  While the $c_k$ are log-concave~\cite{JS}, 
the $x_{k,e}$ may not be.

\subsection{No external field}
In the absence of an external field, we only need to run $\allk$ for $k=0$.  
This represents a huge time savings in comparison to the case $B \not= 0$, 
as then we need to run $\allk$ for all $k \in [0, n/2]$.  
Furthermore, we must choose $S = \emptyset$, which eliminates this step 
from the algorithm.

\subsection{Details} \label{s:details}
The algorithm $\allk$ is really a class of algorithms, each
corresponding to the choice of cycle basis, the order of the 
subgraphs in the basis, the subsets $S$, and the roots $P_S$.  
We briefly discuss these choices here and 
elaborate further in \cite{SM}. 

The choice of cycle basis is central to the performance of $\allk$.  
Experimentally, minimum cycle bases have outperformed fundamental
and random cycle bases in terms of overall speed and variance.  However,
it remains an interesting open problem to determine the optimal basis
for $\allk$.

As for the choice of $S$, we know that $k=0$ implies $S = \emptyset$.
But for $k>0$, we must choose $S$.\footnote{Except if $G$ itself is even, 
in which case there is no choice for $k=n/2$ either.} 
We would like every subset of $V(G)$ of size $2k$ to appear as $S$ at 
least once. However, when $k$ is near $n/4$, $\binom{n}{2k}$ is exponentially 
large in $n$.\footnote{Ideally, we would partition the subsets of $V(G)$ into 
isomorphism classes $\{V_i\}_{i=1}^t$ and choose a representative for class 
$V_i$ to act as $S$ in $|V_i|N/\binom{n}{2k}$ instances.  
However, the number of such classes can also be exponentially large.} 
So instead we are forced to select a reasonable number of such subsets that 
work well in $\allk$.

Once $S = \{v_1, v_2, \dots, v_{2k}\}$ is chosen, we must find $P_S$.  
One such method is to use a spanning tree $T$ of $G$ to 
create $P_S = \bigoplus_{i=1}^k \path{2i-1}{2i}$, 
where $\path{2i-1}{2i}$ is the path from $v_{2i-1}$ to $v_{2i}$ in $T$.

\section{Performance} \label{s:performance}

\subsection{Convergence}\label{s:conv}
In Section \ref{s:phys_quant}, we show how to get unbiased 
estimates of $Z'$ from our unbiased estimates of the $x_{k,e}$.  
To evaluate the efficiency of the algorithm, we need to know how many 
samples ($N$) we need to be reasonably confident about our estimate 
of $Z'$.  The answer depends on the relative variance of our estimate of 
$Z'$.\footnote{By the Central Limit Theorem, we need 
$N = \frac{z_{\delta/2}^2}{\epsilon^2} 
\left(\frac{E[(Z')^2]}{E^2[Z']} -1\right)$ to be within $\epsilon$ 
with probability $1-\delta$, where $z_{\delta/2}$ comes from the normal 
distribution, and $\frac{E[(Z')^2]}{E^2[Z']} -1$ is precisely relative 
variance.} As heuristic sampling methods are relatively new, there are 
not many tools for computing the variance of these algorithms.  
Experimentally, such methods have been shown to work well in practice, 
but a robust theoretical foundation is lacking~\cite{BSSV, Pearl}. 
Therefore, analyzing the variance for this problem remains an important 
open question which deserves further study.

\begin{figure}[!htb]
\centering
\subfloat[]{
\begin{overpic}[scale=.19]{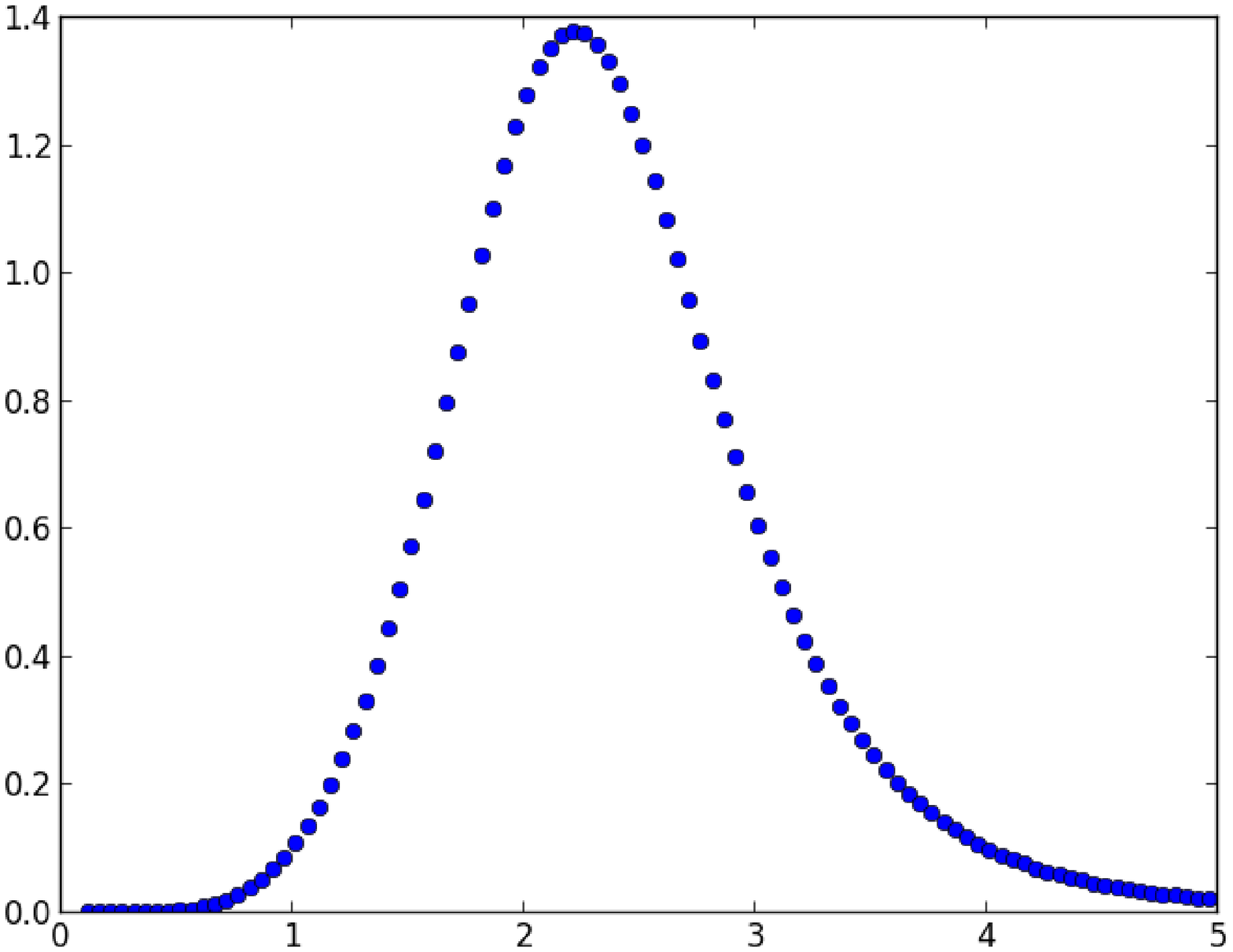}
\put(2,-22){\begin{sideways}\parbox{52mm}{\small{Rel. Var. of $Z'$}}
\end{sideways}}
\put(47,0){$\beta^{-1}$}
\end{overpic}
\label{fig:relvarc0}} 
\subfloat[]{
\begin{overpic}[scale=.19]{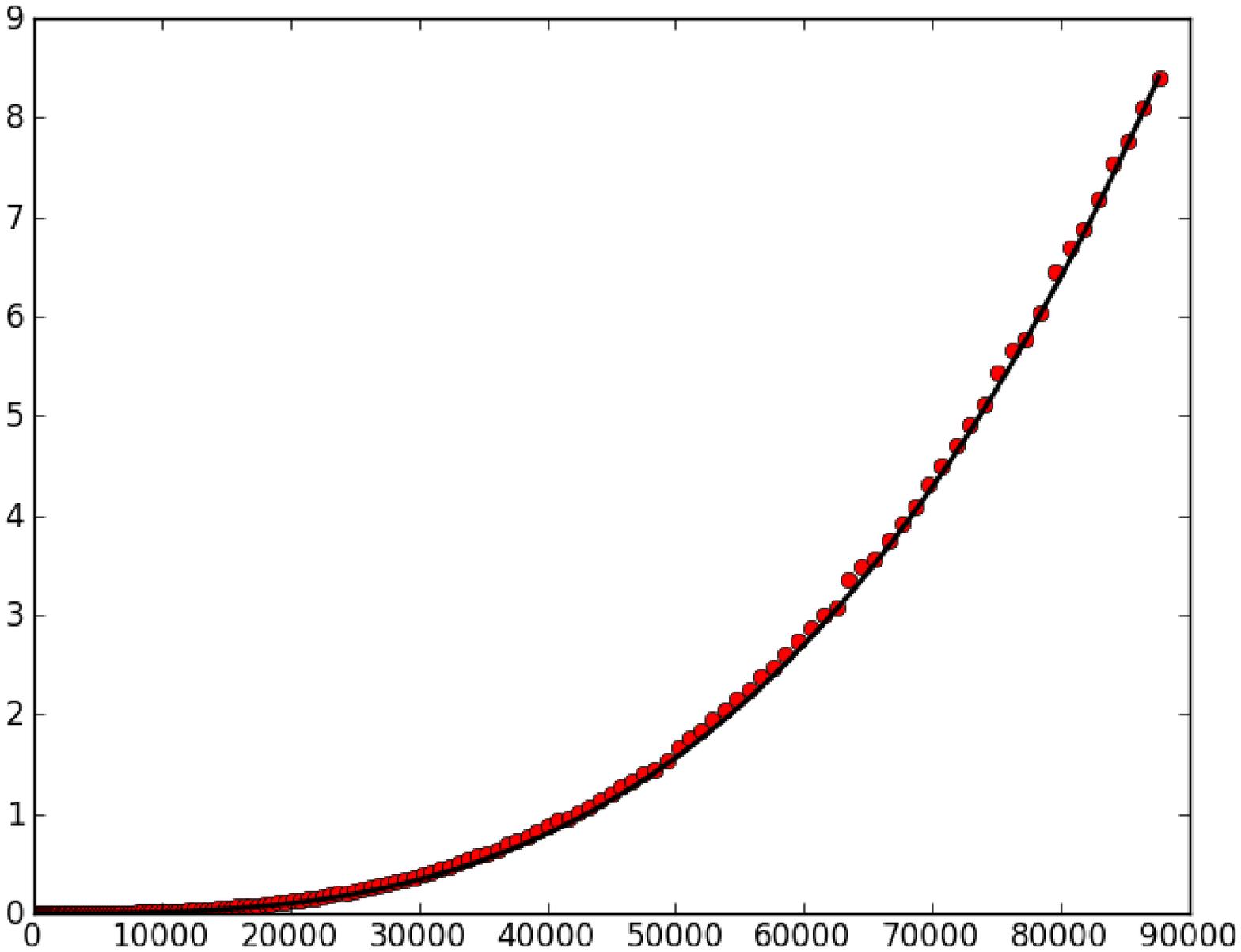}
\put(3,8){\begin{sideways}\parbox{25mm}{Running Time}\end{sideways}}
\put(49,0){$n$}
\end{overpic} \label{fig:times}}
\caption{(Color online)}
\label{}
\end{figure}

\begin{figure}[!htb]
\centering
\begin{overpic}[scale=.45]{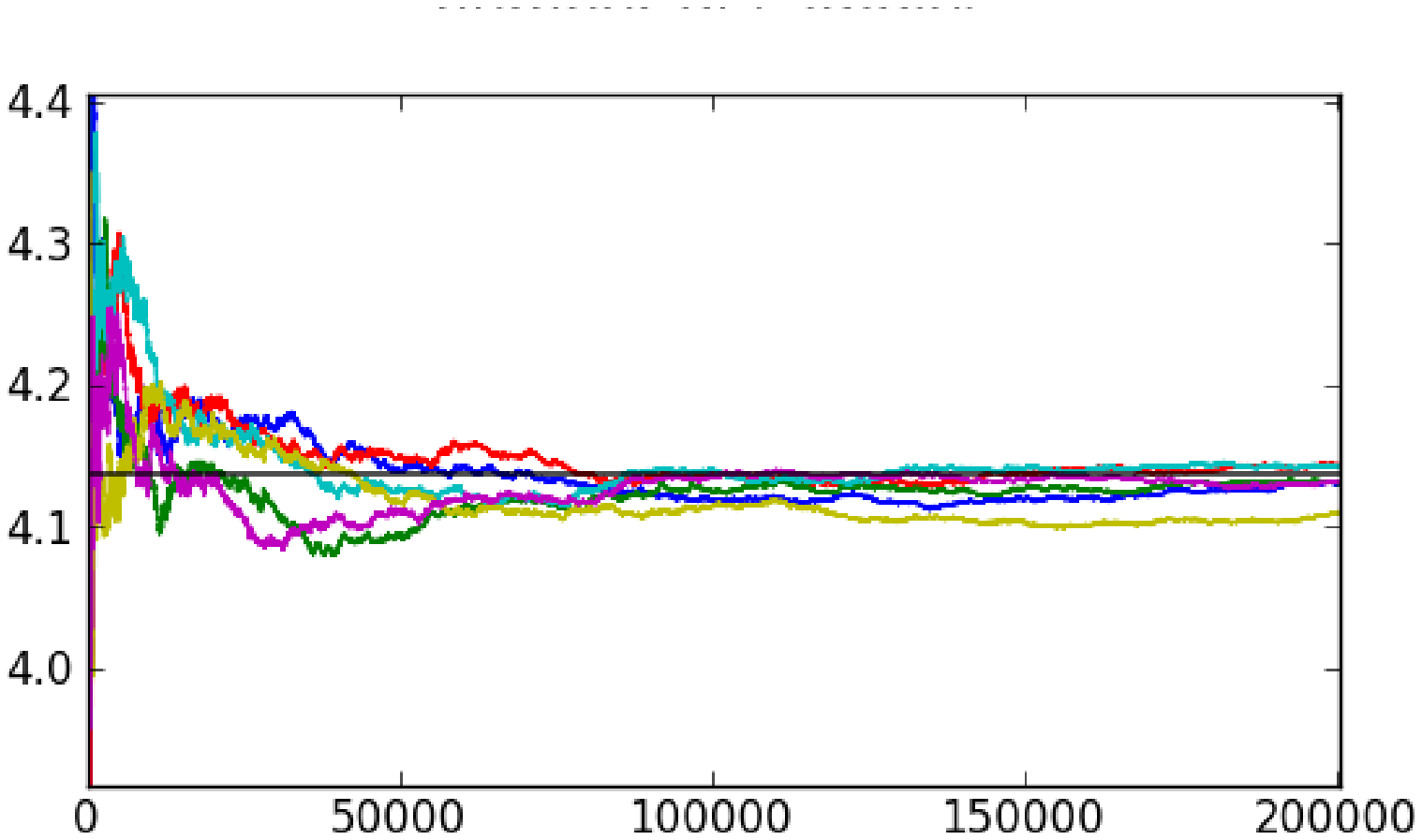}
\put(1,30){$Z'$}
\put(35,1){Number of Samples}
\end{overpic}
\caption{(Color online)}
\label{fig:convergence}
\end{figure}

In our simulations, we find that although $\allk$ is temperature 
independent, the variance is not.  Figure~\ref{fig:relvarc0} shows 
the relative sample variance of our estimate of $Z'$ as a function 
of temperature, for a $4\times 4$ grid with no external field.  
The highest sample variance occurs at the critical temperature, 
$\beta^{-1} \approx 2.269$.  However, even at the critical temperature, 
our estimate of $Z'$ converges quickly. Figure~\ref{fig:convergence} 
presents six separate runs of $\allk$ with $B=0$, and shows the 
convergence to $Z'$ for each run as a function of the number of samples.  
The exact value of $Z'$ is displayed as the straight black line.

\subsection{Running time}\label{s:runningtime}
The number of operations of a single run of Algorithm \ref{alg:Chen} 
as a subroutine of $\allk$ is a function of the number of strata used 
in $\tau(\C,P_S)$ and the number of operations performed
to process each node in $Q_{\current}$.
Recall that our stratifier partitions nodes according to their
level in $\tau(\C,P_S)$ and number of edges.  Clearly, each level 
has at most $m+1$ strata.  Further, there are $m-n+2$ levels.  Hence, 
the number of strata used is at most $(m-n+2) (m+1)$.  For each node 
in $Q_{\current}$, $\allk$ examines its two children, so the total number 
of nodes of $\tau(\C,P_S)$ used by the subroutine is at most
$ 2(m-n+2)m = O(m^2).$ For each of these nodes, we take the symmetric 
difference of two subgraphs and count the number of edges remaining, 
each of which is an $O(m)$ operation.
Thus, each run of Algorithm \ref{alg:Chen} as a subroutine of $\allk$ 
terminates after %a polynomial number of operations, namely 
$O(m^3)$ operations.  For square lattices in dimension $d$, 
the number of operations is $O(d^3n^3)$, as $m=dn$.

\subsection{Implementation}  
We implemented $\allk$ in C, using GMP to deal with the 
large weights generated by the algorithm.  In Figure~\ref{fig:times}, 
we plot our experimental running times for $\sqrt{n} \times \sqrt{n}$ 
grids against the curve $f(n) = 1.25 \cdot 10^{-14} n^3$, 
which matches up well with our bound of $O(n^3)$.  

Typically, one stores graphs as matrices or lists. However, we greatly 
improve the running time of $\allk$ by storing each subgraph $X$ as an 
integer whose bitstring $b_X$ has length $m$; $b_X(e)= 1$ if and only 
if $e\in E(X)$.  Here, $b_{X\oplus Y} = b_X~\text{\sc{xor}}~ b_Y$, so 
taking symmetric differences is quite fast. Further, $|E(X)|$ is simply 
the number of ones in the bitstring, which can also be computed quickly.

One may achieve another increase in speed if 
machine-level instructions for the operation XOR are used
for large integers.  Most modern micro-processors have
such capabilities, as they are used
in scientific computing \cite{intel}.

\section{Physical Quantities}\label{s:phys_quant}
In this section, we show how to use the estimates of the $x_{k,e}$
to calculate physical quantities.  Let $f(X) = f(|\ODD(X)|, 
|E(X)|)$ be any function on subgraphs $X$ which depends only on the 
number of odd vertices and the number of edges of $X$.  We 
can calculate the expected value of $f$ with respect to the distribution
$\pi'(X) =  \lambda^{|E(X)|}\mu^{|\ODD(X)|}/Z'$ from our estimates of 
the $x_{k,e}$ by
\begin{equation}\label{eq:expectation}
\E[f] =\frac{1}{Z'} \sum_{k=0}^{\lfloor n/2 \rfloor} \sum_{e=0}^m f(k,e) 
x_{k,e}\lambda^e \mu^{2k}.
\end{equation}
Notice if $f$ is identically $1$, $Z' \E[f] = Z'$,
and so we can approximate $Z'$, and hence $Z$, by simply looking 
at the double sum. In Theorem~\ref{t:phys_quant}, we show that important 
physical quantities can also be expressed as $\E[f]$ for suitable 
choices of $f$. The proof of Theorem \ref{t:phys_quant} involves taking 
partial derivatives of $\ln Z$ with respect to $\beta$ and $B$ following 
the method of~\cite{JS}. As these calculations are tedious but easy, we 
leave the details to \cite{SM}.

\begin{thm}\label{t:phys_quant}
The mean magnetic moment, mean energy, magnetic susceptibility, and
specific heat can each be written as sums of expectations of random 
variables over the distribution $\pi'$.
\end{thm}

\begin{figure}[!htb]
\centering
\subfloat[]{
\begin{overpic}[scale=.19]{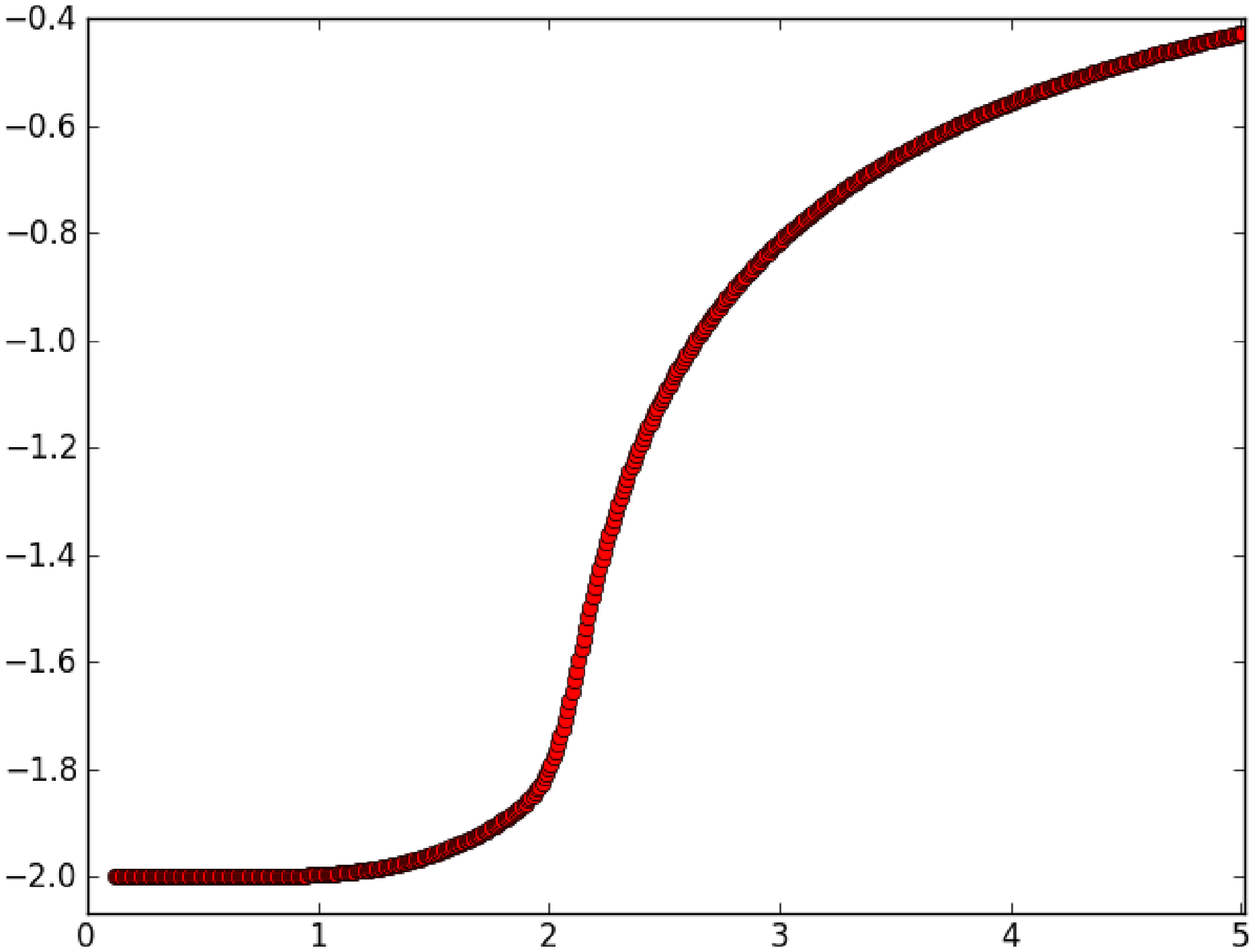}
\put(0,35){\large{$\frac{\energy}{n}$}}
\put(45,0){\scriptsize{$\beta^{-1}$}}
\end{overpic}\label{fig:mean_energy_16}}
%\hspace{.4cm}
\subfloat[]{
\begin{overpic}[scale=.19]{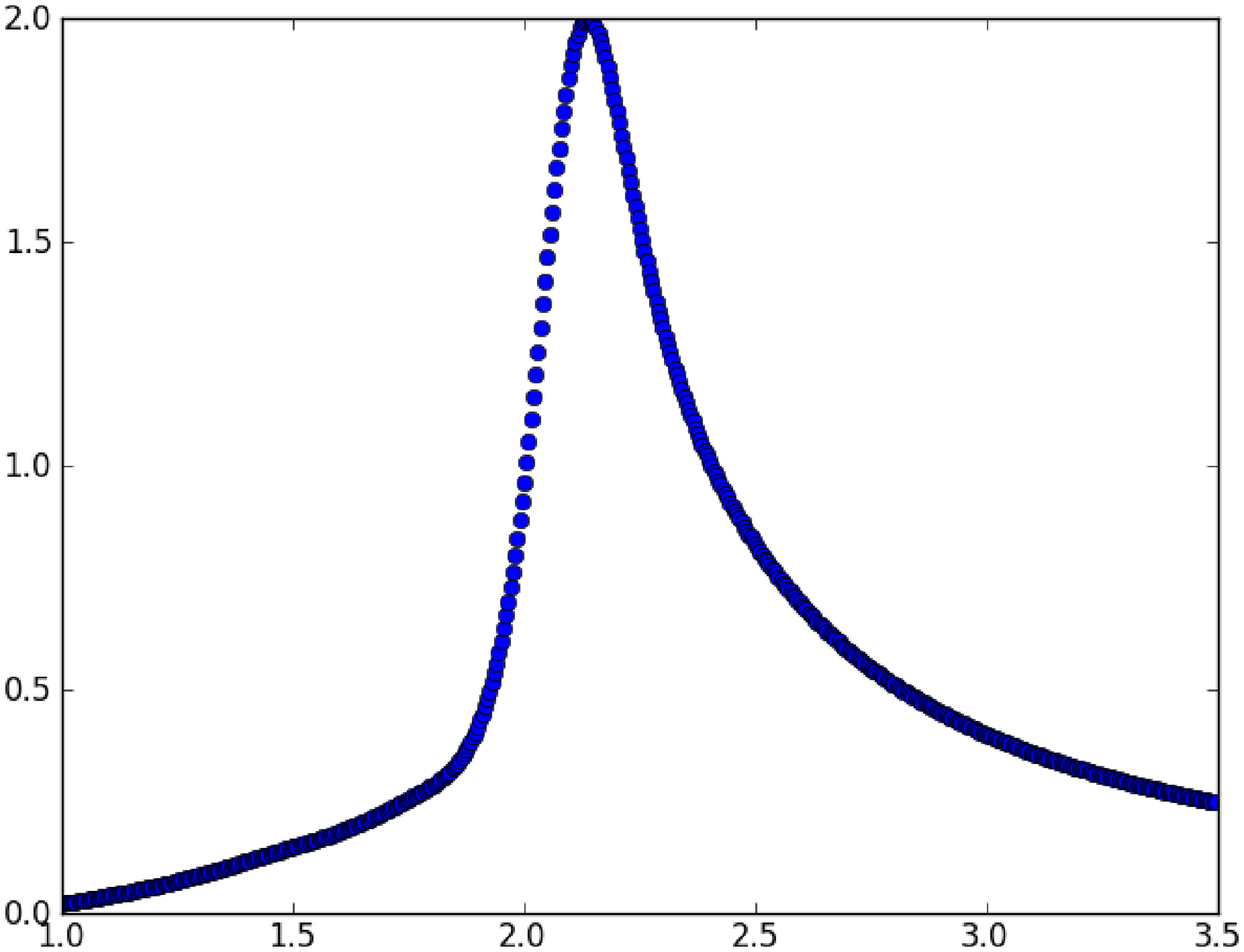}
\put(-5.5,35){$\frac{\mathcal{C}}{nk_{\Boltz}}$}
\put(45,0){\scriptsize{$\beta^{-1}$}}
\end{overpic}\label{fig:specific_heat_16}}
\caption{(Color online)}
\label{fig:mean_energy_and_specific_heat}
\end{figure}

In Figure \ref{fig:mean_energy_and_specific_heat}, we show estimates of 
mean energy and specific heat from $\allk$ with $N = 50,000,000$ on a 
$16 \times 16$ grid as a function of $\beta^{-1}$.  These figures match
those of \cite[p. 252]{gould} nicely.

\section{Conclusions}\label{s:conclusion}
The algorithm $\allk$ is a completely new approach to the problem of
estimating $Z$.  To our knowledge it is the first heuristic sampling
method for this problem.  For this reason, it is difficult to compare
the running time of $\allk$ with the current best-known algorithms,
which are all Markov chain Monte Carlo methods.  What is clear is that
$\allk$ gives us an estimate of $Z$ at \emph{all temperatures
simultaneously} in only $O(m^3)$ operations, where the constant hidden 
by the big-O notation is small.  Bounding the variance of
$\allk$ is an important open problem which is necessary to give a real
understanding of its efficiency.  However, if the goal is to get 
\emph{some} estimate as fast as possible, $\allk$
is an excellent choice. 

Besides analyzing the variance of $\allk$, there are several other
directions for future work.  For example, there are many choices made in
$\allk$ which could be optimized, such as the choice of cycle basis.  These
choices could affect the variance significantly.  One might consider
other tree-search algorithms and compare their performance with that of
$\allk$ and $\bs$.  We also plan to investigate more extensively the 
connections between our heuristic method and MCMC methods.

\begin{acknowledgments}\label{s:acknowledgements}
We wish to thank Professor Ted Einstein at the University of Maryland 
for discussing our ideas and for reminding us of the difference between 
physics and mathematics.
\end{acknowledgments}

\bibliography{Ising}

\end{document}